# Extremely Low Frequency Plasmons in Metallic Microstructures

JB Pendry
The Blackett Laboratory
Imperial College
London SW7 2BZ
UK,

AJ Holden and WJ Stewart
GEC-Marconi Materials Technology Ltd
Caswell
Towcester
Northamptonshire NN12 8EQ
UK

I Youngs
DRA Holton Heath
Poole, Dorset, BH16 6JU
UK

The plasmon is a well established collective excitation of metals in the visible and near UV but at much lower frequencies dissipation destroys all trace of the plasmon and typical Drude behaviour sets in. We propose a mechanism for depression of the plasma frequency into the far infra red or even GHz band: periodic structures build of very thin wires dilute the average concentration of electrons and considerably enhance the effective electron mass through self-inductance. Computations replicate the key features and confirm our analytic theory. The new structure has novel properties not observed before in the GHz band, including some possible impact on superconducting properties.

*Extremely Low Frequency Plasmons in Metallic Microstructures*

Much of the fascination of condensed matter turns on our ability to reduced its apparent complexity and to summarise phenomena in terms of a new 'excitation' that is in fact a composite put together from the elementary building blocks of the material but behaves according to its own simplified dynamics. One of the earliest and most celebrated of these composites occurs in metals and is known as a plasmon [1,2]: a collective oscillation of electron density. In equilibrium the charge on the electron gas is compensated by the background nuclear charge. Displace the gas and a surplus of uncompensated charge is generated at the ends of the specimen, with opposite signs at opposite ends supplying a restoring force resulting in simple harmonic motion,

$$\omega_p^2 = \frac{ne^2}{\varepsilon_0 m_{eff}} \tag{1}$$

The plasma frequency, $\omega_p$, is typically in the ultraviolet region of the spectrum: around 15eV in aluminium.

The plasmons have a profound impact on properties of metals, not least on their interaction with electromagnetic radiation where the plasmon produces a dielectric function of the form,

$$\varepsilon(\omega) = 1 - \frac{\omega_p^2}{\omega(\omega + i\gamma)} \tag{2}$$

which is approximately independent of wave vector, and the parameter $\gamma$ is a damping term representing dissipation of the plasmon's energy into the system. In simple metals $\gamma$ is small relative to $\omega_p$. For aluminium,

$$\omega_p = 15\text{eV}, \quad \gamma = 0.1\text{eV} \tag{3}$$

The significant point about equation (2) is that $\varepsilon$ is essentially negative below the plasma frequency, at least down to frequencies comparable to $\gamma$.

*Extremely Low Frequency Plasmons in Metallic Microstructures*Why is negative epsilon interesting? Cut a metal in half and the two surfaces created will be decorated with surface plasmons [3,4]: collective oscillations bound to the surface whose frequency is given by the condition,

$$\varepsilon_1(\omega_s) + \varepsilon_2(\omega_s) = 0 \qquad (4)$$

Where $\varepsilon_1$ and $\varepsilon_2$ are dielectric functions for material on either side of the interface. Choosing vacuum on one side and metal on the other gives,

$$\omega_s = \omega_p / \sqrt{2} \qquad (5)$$

if we neglect dissipation. It is of course an essential precondition that $\varepsilon$ for the metal be negative. Shape the metal into a sphere and another set of surface modes appear. Two spheres close together generate yet another mode structure. Therefore negative $\varepsilon$ gives rise to a rich variety of electromagnetic structure decorating the surfaces of metals with a complexity controlled by the geometry of the surface.

In fact the electromagnetic response of metals in the visible region and near ultraviolet is dominated by the negative epsilon concept. Ritchie and Howie [5], Echenique [6,7,8], Howie and Walsh [9] and many other researchers have shown how important the concept of the plasmon is in the response of metals to incident charged particles. However at lower frequencies, from the near infra red downwards, dissipation asserts itself and the dielectric function is essentially imaginary. Life becomes rather dull again.

In this letter we show how to manufacture an artificial material in which the effective plasma frequency is depressed by up to six orders of magnitude. The building blocks of our new material are very thin metallic wires of the order of one micron in radius. These wires are to be assembled into a periodic lattice and, although the exact structure probably does not matter, we choose a simple cubic lattice shown below in figure 1.

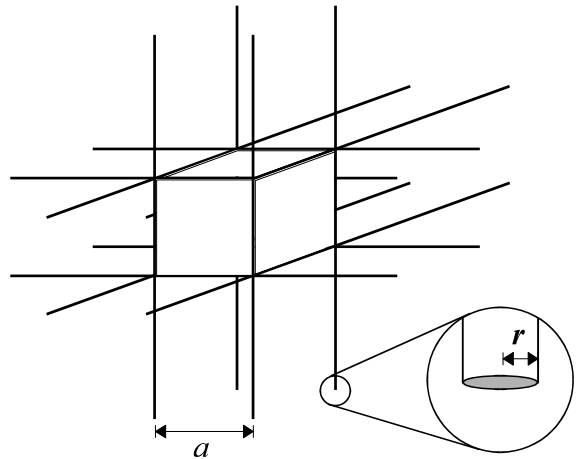

*Figure* 1. The periodic structure is composed of infinite wires arranged in a simple cubic lattice, joined at the corners of the lattice. The large self inductance of a thin wire delays onset of current mimicking the effect of electron mass.

Sievenpiper et al [10] have independently investigated metallic wire structures. Our work differs from theirs in one important respect: we suggest that *very thin wires* are critical to applying the concept of plasmons to these structures.

We now derive the plasma frequency for collective oscillations of electrons in the wires. Consider a displacement of electrons along one of the cubic axes: the active wires will be those directed along that axis. If the density of electrons in these wires is *n*, the density of these active electrons in the structure as a whole is given by the fraction of space occupied by the wire,

$$n_{eff} = n \frac{\pi r^2}{a^2} \tag{6}$$

Before we rush to substitute this number into formula (1) for the plasma frequency we must pause to consider another effect which is at least as important: any restoring force acting on the electrons will not only have to work against the rest mass of the electrons, but also against self-inductance of the wire structure. This effect is not present in the original calculation of the plasma frequency but in our structure it is the dominant effect. It can be represented as a contribution to the electron mass. The important point is that the inductance of a thin wire diverges logarithmically with radius. Suppose a current $I$ flows in the wire creating a magnetic field circling the wire,

$$H(R) = \frac{I}{2\pi R} = \frac{nr^2 ve}{2R} \tag{7}$$

where $R$ is distance from the wire centre. We have also re-expressed the current in terms of electron velocity, $v$, and charge density, $ne$. We write the magnetic field in terms of a vector potential,

$$\mathbf{H}(R) = \mu_0^{-1} \nabla \times \mathbf{A}(R) \tag{8}$$

where,

$$A(R) = \frac{\mu_0 r^2 nve}{2} \ln(a/R) \tag{9}$$

and $a$ is the lattice constant. We note that, from classical mechanics, electrons in a magnetic field have an additional contribution to their momentum of $e\mathbf{A}$, and therefore the momentum per unit length of the wire is,

$$en\mathbf{A}(r) = \frac{\mu_0 e^2 n^2 v}{2\pi} \ln(a/r) = m_{eff} nv \tag{10}$$

where and $m_{eff}$ is the new effective mass of the electrons given by,

$$m_{eff} = \frac{\mu_0 e^2 n}{2\pi} \ln(a/r) \tag{11}$$

This new contribution is dominant for the parameters we have in mind. For instance for aluminium wires,

$$r = 1.0 \times 10^{-6} \text{m}, \quad a = 5 \times 10^{-3} \text{m}, \quad n = 5.675 \times 10^{17} \text{m}^{-3} \text{(aluminium)} \tag{12}$$

gives an effective mass of,

$$m_{eff} = 2.4808 \times 10^{-26} kg$$
$$= 2.7233 \times 10^4 m_e = 14.83 m_p \qquad (13)$$

In other words, by confining electrons to thin wires we have enhanced their mass by four orders of magnitude so that they are now as heavy as nitrogen atoms!

Having both the effective density, $n_{eff}$, and the effective mass, $m_{eff}$, to hand we can substitute into (1),

$$\omega_p^2 = \frac{n_{eff} e^2}{\varepsilon_0 m_{eff}} = \frac{2\pi c_0^2}{a^2 \ln(a/r)} \approx [8.2 GHz]^2 \qquad (14)$$

Here is the reduction in the plasma frequency promised.

Note in passing that although the new reduced plasma frequency can be expressed in terms of electron effective mass and charge, but these microscopy quantities cancel leaving a formula containing only macroscopic parameters of the system: wire radius and lattice spacing. It is possible to formulate this problem entirely in terms of inductance and capacitance of circuit elements. However in doing so we miss the analogy with the microscopic plasmon. Our new reduced frequency plasma oscillation is every bit the quantum phenomenon as is its high frequency brother.

One remaining worry: does electrical resistance in the wires swamp the effect? A more careful calculation including resistance gives the following expression for an effective dielectric function of the structure,

$$\varepsilon_{eff} = 1 - \frac{\omega_p^2}{\omega\left(\omega + \frac{i\varepsilon_0 a^2}{\pi r^2 \sigma}\right)} \qquad (15)$$

where σ is the conductivity of the metal. Typically for aluminium,

$$\sigma = 3.65 \times 10^7 \Omega^{-1} \text{m}^{-1} \quad \text{(aluminium)} \tag{16}$$

and,

$$\varepsilon_{eff} \approx 1 - \frac{\omega_p^2}{\omega(\omega + i \times 0.1\omega_p)} \quad \text{(aluminium)} \tag{17}$$

Thus our new plasmon is about as well defined relative to its resonant frequency as the original plasmon.

To what extent is our theory confirmed by detailed calculations? We have developed a method for calculating dispersion relationships in structured dielectrics [11,12] and we use this to check our analytic predictions. Below in figure 2 we see our calculations of dispersion in our lattice. We choose the most critical case of infinitely conducting lossless wires.

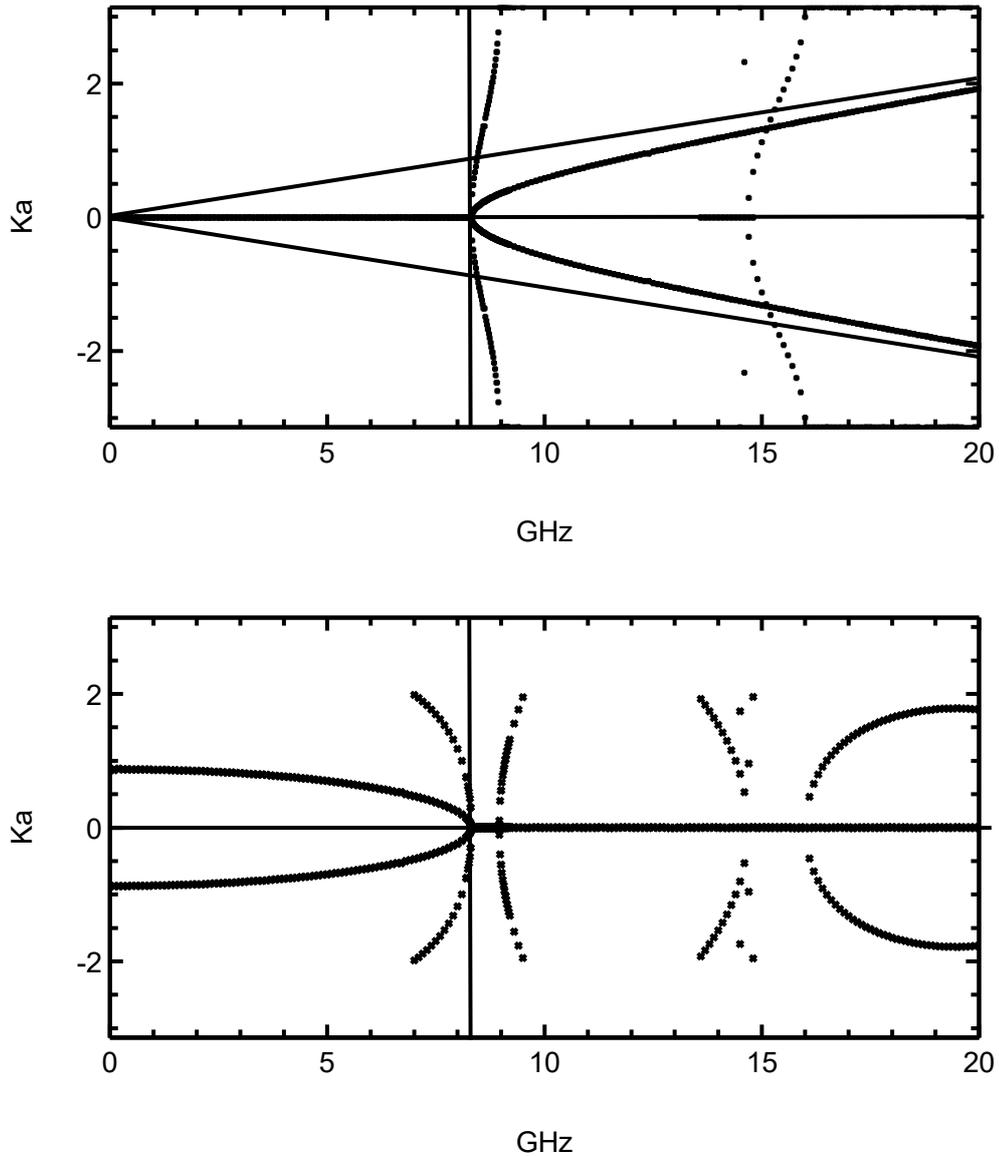

*Figure* 2. Band structure: real (top) and imaginary (bottom) parts of the wave vector for a simple cubic lattice, a=5mm, with wires along each axis consisting of ideal metal wires, assumed 1μ in radius. The wave vector is assumed to be directed along one of the cubic axes. The full lines, largely obscured by the data points, represent the ideal dispersion of the longitudinal and transverse modes defined above assuming a plasma frequency of 8.2GHz. The light cone is drawn in for guidance. Note the two degenerate transverse modes in free space are modified to give two degenerate modes that are real only above the plasma frequency of 8.2GHz.. The new feature in the calculation is the longitudinal mode at the new plasma frequency.

Also shown is the result for dispersion of transverse light obtained by applying our effective dielectric function taken from equation (15) with $\gamma = 0$:

$$K = \omega \frac{\sqrt{\varepsilon_{eff}}}{c_0} = \frac{\sqrt{\omega^2 - \omega_p^2}}{c_0} \tag{18}$$

which gives real bands only above the plasma frequency of 8.2GHz, imaginary bands below. In addition we show the analytic prediction of a dispersionless longitudinal plasmon.

So accurately does our formula reproduce the computed result that the points obscure the analytic line. The computed longitudinal mode agrees very well at $K = 0$ but shows a small degree of dispersion towards the Brillouin zone boundary. Computations for other directions in the Brillouin zone show a similar picture and equation (18) appears to give a good description of the results, at least when $K$ is less than the free space wave vector. It is worth emphasising that at the plasma frequency of 8.2GHz the free space wavelength of light is about 35mm: much greater than the lattice spacing of 5mm. In other words, as far as external electromagnetic radiation is concerned, this structure appears as an effectively homogeneous dielectric medium whose internal structure is only apparent in so far as it dictates $\varepsilon_{eff}$. In this respect it is important that the structure be made of *thin* wires. Equation (14) shows that the function of the small radius is to suppress the plasma frequency. In a thick wire structure in equation (14),

$$\ln(a/r) \approx 1 \tag{19}$$

so that the plasma frequency corresponds to a free space wavelength of approximately twice the lattice spacing. Therefore Bragg diffraction effects would interfere with our simple plasmon picture. Choosing a small radius ensures that diffraction occurs only at much higher frequencies.

In its ideal dissipationless form the structure has the novel feature that below the plasma frequency all electromagnetic modes are excluded from the structure. At sufficiently low frequencies dissipation must take charge in a normal metal, but if superconducting material were employed for the wires and kept well below the transition temperature, dissipation could be small down to zero frequency. In the context of superconductivity it should be noted that plasma frequencies in these structures can be well below the gap energy of a conventional superconductor. Anderson [13] has stressed the role of the plasmon in the electromagnetic properties of superconductors where it appears a 'Higgs Boson' but with a very large mass relative to the superconducting gap. In our new material the Higgs is now well within the gap giving rise to speculation about a more active role for the Higgs in the superconducting mechanism itself. This theme will be pursued elsewhere.

Another allusion to be drawn is to the doping of semiconductors. It is plain from figure 2 that in the GHz frequency range the electromagnetic spectrum is very severely modified. This has been achieved with an extremely small amount of metal: the average density of metal in the structure is less than a part per million: comparable to doping levels in a semiconductor.

The interest in this new material derives from the analogy to be made with the role of the plasmon at optical frequencies. Objects constructed from the new material will support GHz plasmons bound to the surface which can be controlled by the local geometry. Here are possibilities for novel waveguides. Such material is also a very effective band stop/band pass filter. Below the plasma frequency very little can be transmitted; above, and especially in the visible, the structure is transparent.

Another aspect is coupling to charged particles [14]. It is well know from electron microscope studies that metals, metal spheres, and colloids are all efficient at extracting electromagnetic energy from an electron. The mechanism is essentially Cerenkov radiation into the almost dispersionless plasma modes. In our materials the energy scale is much smaller and it is possible to imagine ballistic electrons with a few eV energy injected into our new material where they would have a rather fierce interaction with the low frequency plasmon which could conceivably be exploited in microwave devices.

We have demonstrated that a very simple metallic microstructure comprising a regular array of thin wires exhibits novel electromagnetic properties in the GHz region, analogous to those exhibited by a solid metal in the UV. We trust that the analogy will prove a powerful one and lead to further novel effects and applications.

**Acknowledgements**

This work has been carried out with the support of the Defence research Agency, Holton Heath.

[14]   J.B. Pendry and L. Martín Moreno, Phys. Rev. **B50** 5062 (1994).

Dr Tony Holden & Dr Will Stewart
GEC-Marconi Materials Technology Ltd
Caswell
Towcester
Northamptonshire NN12 8EQ

Dear Tony and Will,

Please find enclosed the first draft of our PRL entitled 'Extremely Low Frequency Plasmons in Metallic Microstructures'. I should be grateful for your comments and suggestions for improvement. What is the routine for getting it cleared for publication?

    with best wishes

                                                John